# Unique Ferroelectric Fatigue Behavior and Exceptional High Temperature Retention in $Al_{0.93}B_{0.07}N$ Films

Wanlin Zhu[1], Fan He[1], John Hayden[1], Quyen Tran[2], Jung In Yang[1], Pannawit Tipsawat[1], Brian Foley[3], Thomas N. Jackson[2], Jon-Paul Maria[1], Susan Trolier-McKinstry[1]

1 Department of Materials Science and Engineering and Materials Research Institute, The Pennsylvania State University

2 School of Electrical Engineering and Computer Science, The Pennsylvania State University

3 Department of Mechanical Engineering, The Pennsylvania State University

**Abstract:**

This paper reports the fatigue and retention behavior for $Al_{1-x}B_xN$ thin films, a member of the novel family of wurtzite ferroelectrics, with an emphasis on the role of capacitor architecture. By modifying the capacitor architecture, and thus thermal and electrical boundary conditions, we create insight regarding the relative importance of intrinsic and extrinsic contributors to the degradation tendencies. Our experiments suggest that bipolar cycling of metal (Pt/W)/$Al_{0.93}B_{0.07}N$/W/$Al_2O_3$ film stacks first induced wake-up, then a region of constant switchable polarization. On additional cycling, the film leakage current increased, and then films underwent dielectric breakdown. For unpatterned first generation $Al_{0.93}B_{0.07}N$ films with 100 nm thick Pt top electrodes survive $\sim 10^4$ bipolar cycles, whereas films with 1000 nm W top electrodes survive $\sim 10^5$ cycles before thermal dielectric breakdown. Sentaurus modeling was used to design an SU8 field plate which improved the performance to $\sim 10^6$ fatigue cycles. It was found that the thermal failures during fatigue were not due to surface flashover events but were associated with hard breakdown events in the dielectric. The films showed excellent retention of the stored polarization state. As expected, data retention was slightly inferior in the opposite state (OS) measurements. However, it is noted that even after $3.6 \times 10^6$ sec (1000 hr). at 200°C, the OS signal margin still exceeded 200 $\mu C/cm^2$. The predicted OS retention is 82% after 10 years baking at 200°C.

## 1. Introduction

Ferroelectric thin films are appealing for non-volatile memories due to the potential for short access times, long-term data storage, large noise margins, and high energy efficiency. It has been shown recently that $Al_{1-x}Sc_xN$,[1,2,3] $Al_{1-x}B_xN$,[4] $AlN$[5] and $Zn_{1-x}Mg_xO$[6] films with the wurtzite crystal structure are all ferroelectric with large switchable polarization that remain stable to high temperatures (the material remains polar to at least 1000°C).[7] This combination of properties is potentially advantageous for ferroelectric random-access memories. However, when developing any new material for ferroelectric memory, understanding the stability of the domain state and the number of times that the polarization can be reversed without loss of switchable polarization is imperative. This paper addresses these factors for a new family of ferroelectrics with the wurtzite structure.

The most widely deployed ferroelectric materials for non-volatile memories are lead zirconate titanate (Pb(Zr$_x$Ti$_{1-x}$O$_3$, PZT) and strontium bismuth tantalate (SrBi$_2$Ta$_2$O$_9$, SBT), with a growing interest in Hf$_{1-x}$Zr$_x$O$_2$ films.[8,9,10] In early development of PZT for non-volatile memories, films with Pt electrodes were susceptible to fatigue, such that the switchable polarization dropped with AC field cycling.[11] There are multiple potential contributions to fatigue, including trapped electronic charges or accumulated oxygen vacancies.[12,13,14] Fatigue has been largely eliminated as a problem in ferroelectric random access memories, either through the introduction of an oxide electrode such as RuO$_2$,[15] IrO$_2$,[16] SrRuO$_3$[17], or La$_{1-x}$Sr$_x$CoO$_3$[18], or by progressive improvements in film quality. In contrast, SBT showed almost no polarization fatigue independent of the metallization employed,[19] but this system has a lower remanent polarization (< 10μC/cm$^2$)[20] than PZT, reducing the memory window. HfO$_2$-based films show robust ferroelectricity[21] with switching endurance up to 10$^{10}$ cycles.[22] It has been reported in this system as well that fatigue may depend on migration of oxygen vacancies.[23] Notably, during fatigue, PZT often undergoes soft failure, where polarization is lost without destroying the film, whereas HfO$_2$-based ferroelectrics are more susceptible to hard breakdown events.

Retention, which quantifies the ability of a ferroelectric to retain the stored polarization state, is another critical property for non-volatile memories. Loss of data can occur due to aging of the written state, or to the progressive development of global or local imprint.[24,25] Any factor that produces a local change in the stability of the domain state, including injected charges,[26] asymmetric electrodes, strain,[27] grain boundaries,[28,29] thermodynamically favored domain orientations[22], or defects[30,31] can change the retention characteristics. Retention is quantified in terms of both same state and opposite state retention; typically, the usable polarization drops linearly with the logarithm of wait time. In the case of sputtered PZT films, same state retention characteristics tend to be more stable than opposite state retention, due to imprinting of the stored state.[32] Data loss accelerates as the temperature is increased during the wait time between read and write due to thermal activation of the depolarization. Rodriguez et al. reported that in integrated memories with 70 nm PZT films as the storage element, the activation energy associated with polarization loss was 0.28 eV.[8] Furthermore, the retention did not change substantively, even after 10$^{12}$ switching cycles. Likewise, long-term data retention has been reported for epitaxial Hf$_{0.5}$Zr$_{0.5}$O$_2$ films.[33,34] Data of this type is essential to viable memory devices.

Fatigue and retention loss are largely unreported in nitride wurtzite ferroelectrics. Therefore, this paper discusses the mechanisms for polarization loss due both to bipolar cycling of the electrical field, and temperature exposure. The first generation of these films showed robust information retention in both same state and opposite state, even up to 200°C. However, they were susceptible to hard breakdown events in bipolar cycling.

## 2. Results and Discussion

*Bipolar fatigue*

The wurtzite nitrides differ in a number of important ways from perovskite ferroelectrics used in memory applications, as is illustrated in Fig. 1, a hysteresis loop comparison between a lead zirconate titanate (PZT) 30/70 thin film and AlN with 7% B substituted for Al. In investigating the origins of fatigue in ferroelectric nitrides, it is important to recognize: **i.** the margin separating coercive and breakdown fields are modest (perhaps 15% at room temperature in $Al_{1-x}B_xN$) and the required field magnitudes are high, typically between 2 and 6 MV/cm; **ii.** ferroelectric nitrides switch ~ 3-5 × more polarization per field cycle than PZT depending on composition and orientation; **iii.** The polarization – electric field hysteresis loops are square leading to substantially higher instantaneous switching currents; **iv.** these large currents exceed the limits of conventional instrumentation, due to the high instantaneous powers needed and **v.** $Al_{1-x}Sc_xN$ and $Al_{1-x}B_xN$ early in their development cycle.

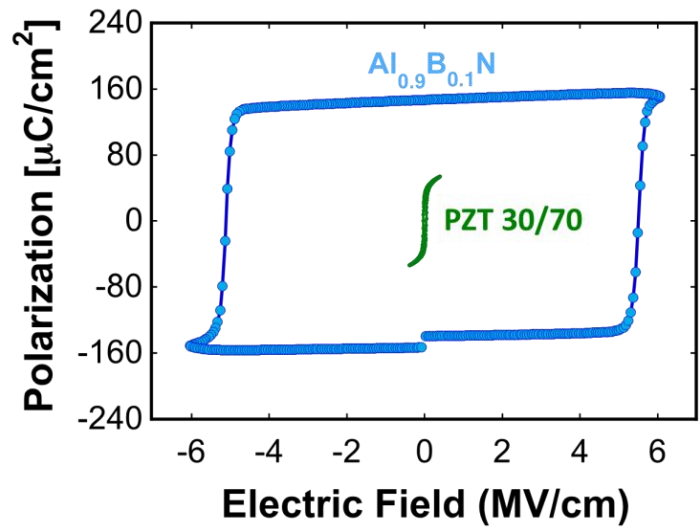

Fig. 1: Polarization hysteresis comparison for a PZT (30/70) thin film and an aluminum boron nitride thin film. The dissipated energy for AlBN is > 400 times greater than PZT based on loop areas

The capacitors in this study are prepared on *c*-plane sapphire substrates with W bottom electrodes and an ~ 300 nm thick $Al_{0.93}B_{0.07}N$ dielectric layer, both prepared by magnetron sputtering at 400 °C. Top electrodes were prepared *ex situ* using a variety of geometries so as to explore the fatigue process. Fig. 2 shows trends in apparent remanent polarization vs field cycling for a collection $Al_{0.93}B_{0.07}N$ capacitor collection. The structures include (a) an AlBN layer with a 100 nm thick, dot-shaped Pt top electrode with unpatterned ferroelectric and bottom electrode layers, (b) an identical $Al_{0.93}B_{0.07}N$ -bottom electrode stack with a 1 µm thick W dot shaped top electrode, and (c) the same $Al_{0.93}B_{0.07}N$ dielectric-bottom electrode stack with a 100 nm thick and a thick photoresist (SU8) field plate. These structures provide different thermal and electrical boundary conditions.

The simplest structure (thin isolated top electrode) is shown in Fig. 2(a). During the initial ~ 400 cycles, the polarization gradually increases following the wake-up phenomenon that was previously reported by Zhu *et al.*, who noted that the number of cycles required can be a function of frequency and current limits during cycling – longer field cycles and higher currents produce faster wake-up.[35]

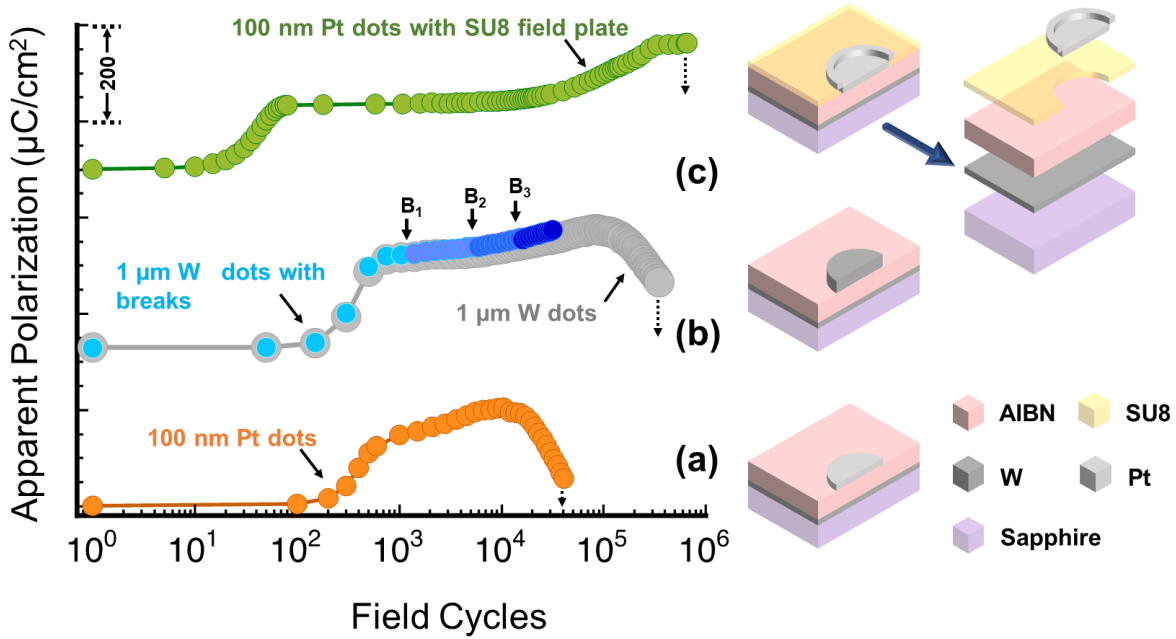

Fig. 2: Evolution of the apparent remanent polarization, "$P_r$" with cycle number on (a) $Al_{0.93}B_{0.07}N$ with 100 nm Pt top electrode; (b) $Al_{0.93}B_{0.07}N$ with 1000 nm W top electrode and (c) $Al_{0.93}B_{0.07}N$ -SU8 film with 100 nm Pt top electrode

Additional cycles from ~500 to ~10,000 are accompanied by an increasing apparent polarization. The increase in the apparent polarization is due to addition of a substantial leakage current contribution, associated with some form of material or interfacial degradation that reduces insulation resistance. Between ~10,000 and ~40,000 cycles the apparent polarization falls monotonically until the capacitor finally fails to an open circuit condition. This type of failure occurs primarily when the voltage amplifier is set in low-power mode (with a current limitation of 20mA), and it is accompanied by an electrode failure process where concentric rings of electrode material and dielectric are lost but the integrity of the remaining interior electrode area persists. A "tree-ring" like surface morphology (as shown in Fig. S3(a)) emerges, until at some point, a catastrophic event occurs. Thus, the apparent polarization loss results from a decreasing capacitor area, rather than a fatigue process that limits switching of the ferroelectric phase.

In measurements in which a higher current limit is set for the power supply (400mA) failure often occurs in fewer cycles following a small number of thermal failure events located in the capacitor electrode interior; in this case, the sample does not produce a "tree-ring" pattern of thermal failure events upon field cycling. A representative sample that failed in this fashion is shown in Fig. S3 (b). As is discussed in detail later, it is likely that tree-ring like patterns are promoted by field concentrations at the electrode periphery, while interior thermal failure events are associated with material or morphological defects in the film or bottom electrode.

One possible origin of the increase in apparent polarization on cycling as shown in Fig. 2(a) would be that thermal heating from polarization reversal. To test this possibility, a second set of capacitors with similar dielectric layers were prepared with one-micron-thick W top electrodes that reduce the electrode resistivity, and hence electrode contributions to Joule heating. These capacitors were tested in two ways: continuous field cycling identical to that used in Fig. 2(a); and a field cycling process including one-hour rest periods during which thermal excursions – if present – will dissipate. Both results are shown in Fig. 2(b). There are two outcomes of importance to note: **i.** thicker electrodes extended the region where apparent polarization increases with field cycling and the number of cycles to hard failure – both by ~ one order of magnitude; and **ii.** Incorporating breaks in the cycling process does not appreciably change the fatigue characteristics. As shown in Fig. 2, higher leakage currents were retained even after the wait states. Thus, the rise in leakage currents associated with bipolar cycling is not due to a gradual temperature rise induced by switching. Instead, it is observed that the data picked up at values similar to those prior to the break in cycling.. These experimental results are consistent with a 2D model demonstrating that the steady-state temperature increase should be under 1°C for these testing conditions; the details are provided in the supplemental information.

The tree-ring morphology and concentration of degradation events near the electrode perimeter suggests field concentrations at the electrode periphery are substantial. Fig. 3 shows Sentaurus simulations of the electric field profiles across the capacitor and through its thickness for an $Al_{0.93}B_{0.07}N$ dielectric and an SU8 field plate. The left peak shows the electric field at the electrode perimeter of the top electrode directly on the ferroelectric. The maximum electric field is enhanced by approximately a factor of two to three relative to the bulk capacitor near the electrode edge, as is expected. This high field is very likely to be the source of dielectric breakdown events that develop prior to true ferroelectric fatigue. The right-hand simulation is for the field plate design, in which the top electrode extends over the SU8 dielectric. Because the dielectric is thick and has a low relative dielectric constant (0.7 µm and 3, respectively) the simulation shows that most of the electric field reduction occurs in the vertical electrode region. The field plate reduces the field concentration factor by ~ 20%.

To test whether this design improves the lifetime experimentally, a third set of capacitors were fabricated with a field plate design that partially mitigates the concentration effect. As shown in Fig. 2(c), for these capacitors, there is an initial wake-up period, followed by an apparent polarization plateau between ~100 and ~ 20,000 cycles where the remanent value is constant. Between ~20,000 and ~300,000 cycles, the apparent polarization increases gradually, and again finds a polarization plateau until total failure (open circuit) at ~500,000 cycles. This second plateau is associated with the current limitations of the measurement electronics. Importantly, the gradual polarization loss before total failure is no longer present. The initial apparent polarization plateau with cycling indicates that the field plate effectively reduces perimeter concentrations thus revealing a service condition with no accruing damage that adds leakage contributions to calculated polarization – even though the reduction is modest. This condition, however, is temporary and yields to a region where damage accumulates, albeit more slowly than if the field plate were removed until ultimate failure. Images of the failed capacitors with field plate reveal failure events primarily in the capacitor interior that are possibly associated with material or morphological inhomogeneities. the cycling performance was increased by approximately and order of magnitude to $10^6$ cycles. It also suggests that electrode designs which further reduce field concentrations will likely produce significantly improve cycling performance.

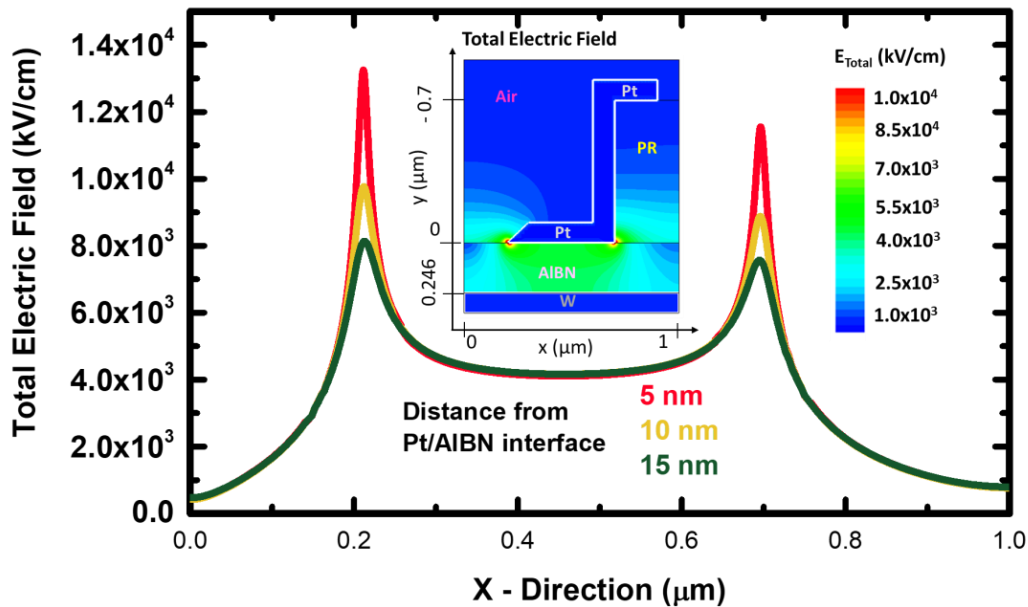

Fig. 3: 2-D Electric field simulation for an $Al_{0.93}B_{0.07}N$ capacitor where the left edge is surrounded by air while the right edge is surrounded by SU8 photoresist (*i.e.,* the field plate material); three line-scans along the x axis represent the electric field value at the plane 5 nm, 10 nm, and 15 nm away from Pt/$Al_{0.93}B_{0.07}N$ interface into the film, respectively.

These observations suggest that $Al_{0.93}B_{0.07}N$, and likely other ferroelectric nitrides, are susceptible to "extrinsic" fatigue mechanisms associated with electrodes and electrode-dielectric interfaces due to the large coercive fields, large instantaneous switching currents, and large dissipated energies per switching cycle. Consequently, factors associated with capacitor design including shape, perimeter-to-volume ratio, thermal boundary conditions, and dielectric discontinuities can dominate the fatigue process, *i.e.,* failure from extrinsic mechanism may occur prior to any intrinsic reduction of switchable polarization from locked domains. Accessing and understanding the material fatigue will likely require more sophisticated geometries which reduce field concentrations. The present demonstration that adding a simple low-permittivity field plate strongly alters the failure path suggests opportunities for substantial improvement.

In summary, these Al$_{0.93}$B$_{0.07}$N films did not show significant reduction in polarization before catastrophic failure upon bipolar cycling. Failed films were marked by the appearance of thermal failure events in which the film and top electrode were locally vaporized, leaving behind a crater. Thick top electrodes and field plate increased the fatigue lifetime of Al$_{0.93}$B$_{0.07}$N films, due to reducing electrode edge electric field concentration. It is speculated that if weak spots in the films can be reduced by optimized deposition conditions, and/or electrodes with larger Schottky barrier heights are utilized, then the fatigue lifetime of these films will further rise.

*Retention*

A second property which is essential in nonvolatile memory is retention of the data over time. Retention measurements were conducted with bake temperatures of room temperature, 100°C, 150°C, or 200°C. Fig. 4 (a) and (b) shows the resulting data for same state and opposite state retention. As was previously reported, the remanent polarization in these films is a very weak function of temperature, such that the initial polarization exceeds 120 µC/cm$^2$ over this temperature range.[5] Despite these very large polarizations, the polarization loss is negligible after 500 h at 100°C and 150°C. Even at 200°C, the signal margins exceed 200 µC/cm$^2$ after bake times of 3.6×10$^6$ sec (1000 hours). In contrast, scaled PZT ferroelectric random-access memory (FeRAM) shows signal margins under 10 µC/cm$^2$ after comparable heat treatments.[8] This suggests that these Al$_{0.93}$B$_{0.07}$N films are excellent candidates for non-volatile memories, with data retention times at high temperatures even in very early generation films.

The activation energy ($E_{a\Delta P_1}$) for opposite state (OS) margin reduction after baking for 1 hour and the polarization loss rate (m) in retention measurement were calculated using Eq. (1) and (2).[8]

$$P(t,T) = P_0(T) - \Delta P_1(T) - m(T)\ln(t) \tag{1}$$

$$P(t,T) = P_0(T) - A_{\Delta P_1} exp\left(\frac{-E_{a\Delta P_1}}{k_B T}\right) - m(T)\ln(t) \tag{2}$$

$P(t,T)$: polarization margin after bake at T (temperature, K) for t (time, hour).

$P_0(T)$: initial polarization margin before bake.

$\Delta P_1(T)$: margin reduction after 1 hour of bake at T.

$A_{\Delta P_1}$: preexponential constant.

$k_B$: Boltzmann's constant.

As shown in Fig. 5, according to Eq. (2), an Arrhenius plot produces the activation energy $E_a$ for polarization loss $\Delta P_1(T)$ of 0.10±0.03eV in the range from 25°C to 200°C, smaller than the activation energy reported for PZT films. The underlying physics of this polarization loss is currently under study. The parameter that describes the *rate of* polarization loss, *m*, as described in Eq. (1), has an activation energy of 0.21±0.01eV. This activation energy is approximately 7.5 times larger than thermal activation of subcoercive field domain wall motion (~28meV) at similar temperatures [5]. The OS margin as showed in Fig. 4 (b) can still keep up to 82% of the pristine value after 10 year with 150oC baking, based on calculation with Eq (1).

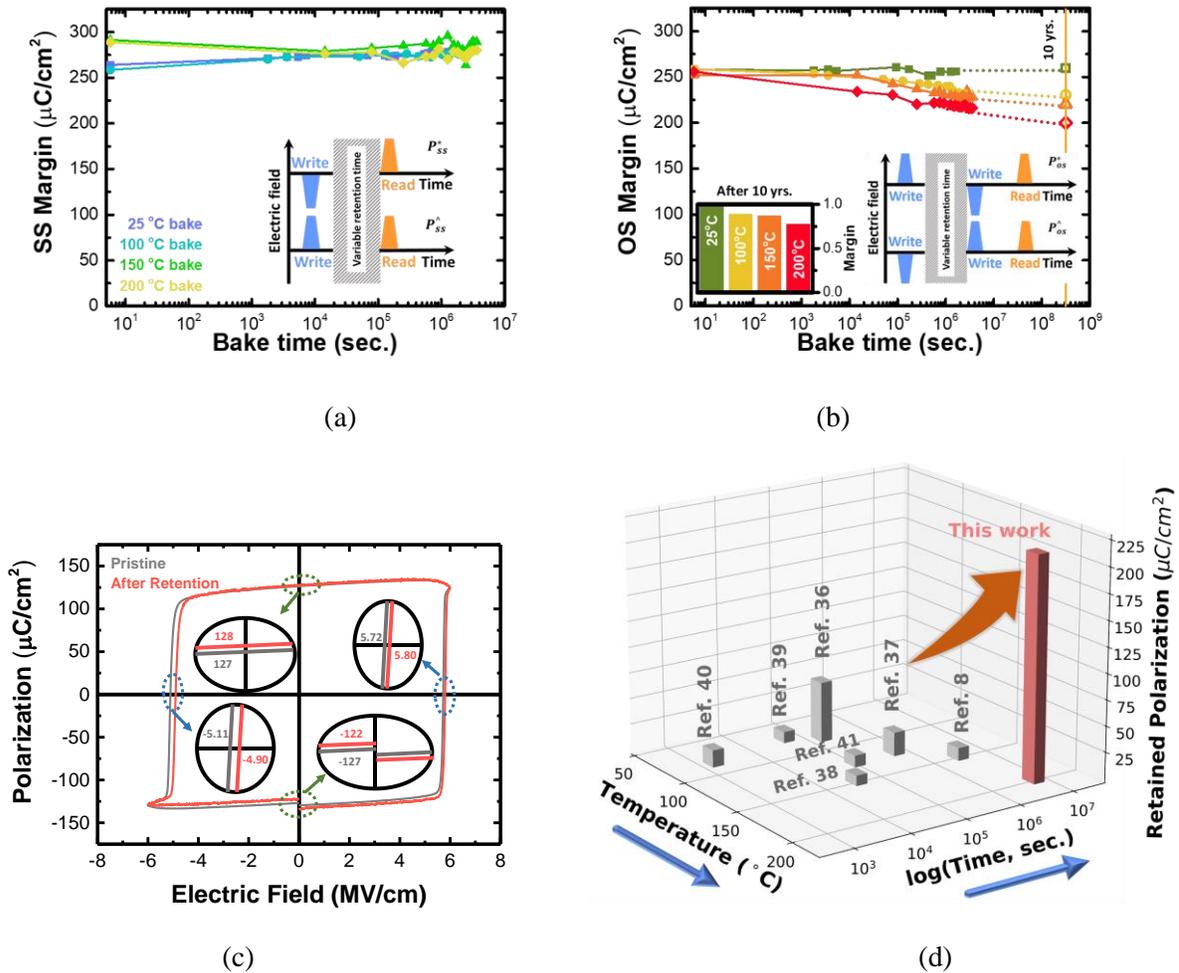

Fig. 4: Retention data for $Al_{0.93}B_{0.07}N$ films. (a) same state polarization margin and (b) opposite state polarization margin measured at room temperature, 100°C, 150°C and 200°C as a function of bake time; (c) Polarization – Electric field loops of a $Al_{0.93}B_{0.07}N$ films measured in the pristine state and after 3.6×10⁶ sec. (1000 hr) baking at 150°C (grey line and text represent pristine state result, while the red line and text are the state after the bake); (d) retention data comparison[8,36-41]

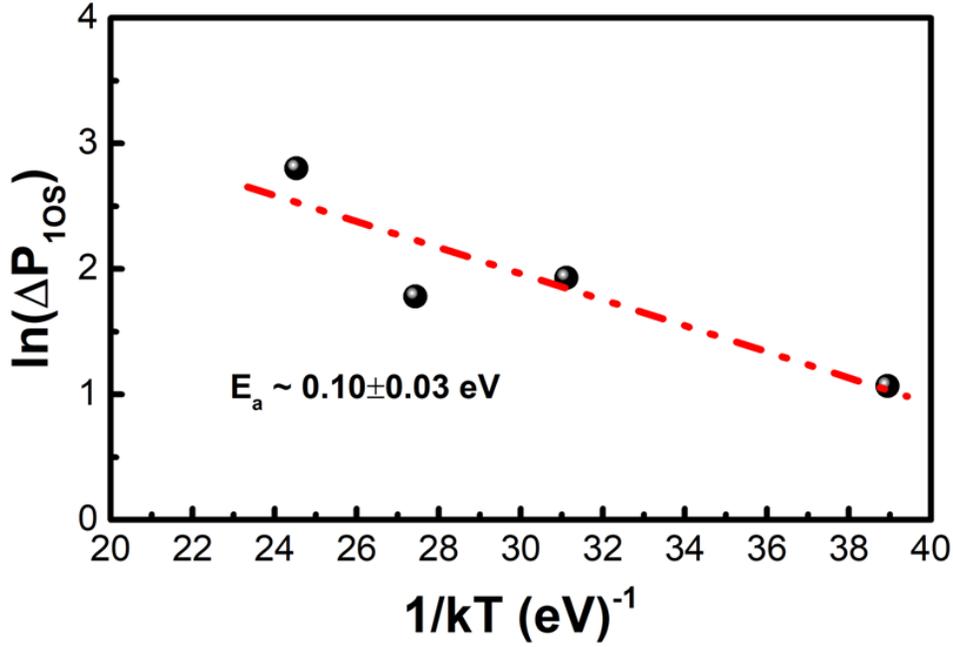

Fig. 5: Temperature dependence of opposite state $\Delta P_1$ between 25°C and 200°C

Polarization – electric field hysteresis loops were also checked before and after retention measurement, as shown in Fig. 4 (c). In the pristine state, the coercive fields $E_c$ were -5.11 and 5.72 MV/cm, and the remanent polarization $P_r$ was ±127 µC/cm². After baking at 150°C for 3.6×10⁶ sec. (1000 hours), the $E_c$ were -4.90 and 5.80 MV/cm, $P_r$ was -122 and +128 µC/cm². Thus, there were only small changes in the imprint and the magnitude of the polarization. There is a small polarization drop; the very slight positive shift implied downward (top to bottom electrode) imprint, consistent with the preferred downward polarization direction of the film as grown. Comparing with previously reported retention data[8,36,37,38,39,40,41], as shown in Fig 4 (d), this work showed that $Al_{0.93}B_{0.07}N$ shows excellent retention performance. Some short retention data on Sc-doped AlN film has been reported recently[42], which could not compare with our work, due to the completely different voltage profile and retained polarization calculation.

While the films here are thick, recent data suggests that ferroelectricity is retained in AlN-based films down to at least 9 nm,[43] albeit most group report increasing leakage currents in thinner films.[44,45,46] Moving forward, it will be essential to demonstrate scaling of the $Al_{1-x}B_xN$ films to smaller film thicknesses without degrading their insulation resistance.

## 3. Conclusions

The fatigue and retention characteristics of $Al_{0.93}B_{0.07}N$ wurtzite-structured ferroelectric thin films were investigated. The films lifetime for bipolar cycling showed approximately two orders of magnitude increase on adding an SU8 field plate for the circular top electrode. It was found that the films showed excellent retention of stored information, with signal margins in excess of 200 µC/cm$^2$ after $3.6\times10^6$ sec. (1000 hr) baking at 200°C for metal-ferroelectric-metal stacks. There is little change in the imprint of the loops even after such extended baking times. The activation energies for polarization loss were 0.10±0.03eV for $P_1$, and 0.2±0.01eV for parameter $m$. On bipolar cycling of the hysteresis loops, the films were found to be susceptible to dielectric breakdown events that first led to an increase in leakage currents, and subsequently to localized failure events in the films. These events were exacerbated in the presence of local field concentrations near the electrode perimeter. It is anticipated that further decreasing the ratio of the coercive field to the breakdown field will allow increased reliability in these wurtzite ferroelectrics.

## 4. Experimental Procedure

$Al_{0.93}B_{0.07}N$ films were grown on $W/Al_2O_3$ substrates, as described elsewhere.[4] In brief, films of approximately 300 nm thickness were reactively sputtered using Al (pulsed DC power supply) and B (RF power supply) targets.

Considering the low permittivity of the $Al_{0.93}B_{0.07}N$ film, electric field concentration was expected caused by field curvature along the electrode edge. To find an optimized field plate for this film, Sentaurus Device was used to simulate the device structure. Because the electrodes were patterned by lift-off, the electrode was modelled with an angled cross-section. A radius of curvature of 5 nm was used at the electrode edge to reduce simulation singularity effects.

Four sets of top electrodes were explored: 100 nm thick Pt electrodes defined by a single layer lift-off process, 1 µm W electrodes that were deposited through a shadow-mask after the ferroelectric film deposition, a guard ring design and 100 nm thick, $\phi$220 µm circular Pt electrodes on an SU-8 insulation layer in a field plate design. For the last geometry a Heidelberg MLA 150 direct writing lithography tool was used. No significant improvement was noted using the guard ring, so those results are not discussed further in the manuscript.

Fatigue measurements were carried out by continuously cycling the film at 100Hz with a 5-6 MV/cm triangular AC field utilizing a custom Sawyer-Tower circuit. Retention measurements were carried out with a Multiferroics Precision System (Radiant, Albuquerque) with bake temperatures at room temperature, 100°C, 150°C and 200°C on a 245 nm thick $Al_{0.93}B_{0.07}N$ film on W/sapphire. The samples were exposed to PUND measurements using a pulse width of 1ms and a pulse delay of 10 ms at an electrical field of 6 MV/cm to wake up the polarization switching in the film. The same state and opposite state retention were then measured as described by Yoon et al.[47], using a Radiant Technologies Precision Multiferroic Analyzer, with 1 ms pulse width, 10 ms pulse delay, and an electric field of 5-6 MV/cm. Schematics of the voltage profiles employed are shown in Fig. S1 in the supplemental materials and Fig. 4 (a), (b). The polarization difference, $\Delta P$, was defined as

$$\Delta P = P^* - P^{\wedge}$$

where $P^*$ is the switched polarization and $P^{\wedge}$ is the non-switched polarization.

## Acknowledgements

The authors gratefully acknowledge funding from Defense Advanced Research Projects Agency (DARPA) through the Tunable Ferroelectric Nitrides (TUFEN), program grants HR0011-20-9-0047 and W911NF-20-2-0274 for identification of the optimal synthesis conditions and the fatigue and retention measurements. Preparation of the ferroelectric films used for this studies, analysis of the retention data and the modeling of the field distributions were supported by the U.S. Department of Energy, Office of Science, Office of Basic Energy Sciences Energy Frontier Research Centers program under Award Number DE-SC0021118. The authors would also like to thank Prof. Geoff Brennecka of the Colorado School of Mines for helpful discussions.